\documentclass[english,onecolumn,prx,preprintnumbers,amsmath,amssymb,superscriptaddress]{revtex4-2}
\usepackage{verbatim}
\usepackage{graphicx}
\usepackage{amssymb}
\usepackage{ulem}
\usepackage{xcolor}

\usepackage{babel}
\usepackage{float}
\usepackage{titlesec}
\makeatletter

\def\maketitle{
	\@author@finish
	\title@column\titleblock@produce
	\suppressfloats[t]}
\makeatother

\makeatother

\begin{document}
\preprint{XXX}

\newcommand{\Vac}{V_{\mathrm{ac}}}
\newcommand{\Vdc}{V_{\mathrm{dc}}}
\newcommand{\omd}{\omega_d}

\title{Tunable Nonlinear Landscapes in Graphene Nanoelectromechanical Systems}

\author{Ateesh K. Rathi}
\affiliation{Department of Physics, Indian Institute of Technology - Kanpur, UP-208016, India}

\author{Rajan Singh}
\affiliation{Department of Physics, Indian Institute of Technology - Kanpur, UP-208016, India}

\author{Javed A. Mondal}
\affiliation{Department of Physics, Indian Institute of Technology - Kanpur, UP-208016, India}
\author{Arnab Sarkar}
\affiliation{Department of Physics, Indian Institute of Technology - Kanpur, UP-208016, India}

\author{Ryan J.T. Nicholl}
\affiliation{Department of Physics and Astronomy, Vanderbilt University, Nashville, Tennessee 37235, USA}	

\author{Kirill I. Bolotin}
\affiliation{Department of Physics, Freie Universitat Berlin, Arnimallee 14, Berlin 14195, Germany}

\author{Saikat Ghosh}
\email{gsaikat@iitk.ac.in}
\affiliation{Department of Physics, Indian Institute of Technology - Kanpur, UP-208016, India}

%\author{Sagar Chakraborty}

%\keywords{graphene, nanoelectromechanical systems, internal resonance, phononic frequency comb, high-harmonic generation, period doubling}

\date{\today}

\begin{abstract}
Nonlinear nanomechanical resonators give convenient solid-state access to classical analogs of extreme nonlinear optics and to phononic signal processing. Here we report integer high-harmonic generation and phononic frequency combs in a suspended monolayer graphene drum. A gate voltage breaks the out-of-plane symmetry of the membrane and tunes its fundamental flexural mode onto a 1:2 internal resonance with a higher mode at twice the frequency, where the quadratic coupling between the two modes becomes large. A single drive tone then generates phase-locked integer harmonics in sequence, and at larger drive these fill in to a dense frequency comb. Raising the drive further, we find a reverse period-doubling transition: the comb spacing doubles, the line density halves, and energy flows back into the even-order comb lines. The measured spectra yield the quadratic ($\zeta$) and cubic ($\beta$) nonlinear coefficients of the membrane. These results show how the tunable nonlinear landscape of graphene supports distinct dynamical regimes on demand, allowing a single gated device to act in turn as a frequency multiplier, a broadband comb source, and a chaotic generator.
\end{abstract}
\maketitle

\section*{Introduction}

In nanoelectromechanical systems (NEMS), nonlinearity used to be something to design around; it is now something to put to work.~\cite{doi:https://doi.org/10.1002/9783527626359.ch1,RevModPhys.94.045005} Operated deliberately in the nonlinear regime, these devices perform signal processing,\cite{Tadokoro2018} frequency metrology,\cite{Maillet2018} and forms of sensing and computation with no linear analog.\cite{Weber2016,PhysRevApplied.19.054074} Suspended two-dimensional materials, and monolayer graphene in particular, are attractive here:\cite{Davidovikj2017,Fan2024} their low mass, atomic thickness, and high tensile strength make intrinsic nonlinearities unusually strong,\cite{Chen} and their compliance lets a gate voltage reshape the potential in which they move.\cite{PhysRevLett.124.046101} Graphene resonators have in this way been pushed to show parametric amplification,\cite{Singh2018} phonon lasing,\cite{PhysRevLett.110.127202} and phononic frequency combs (PFCs),\cite{PhysRevLett.118.033903,Singh2020,Ye2025,doi:10.1126/sciadv.adv9984,Sarkar2026} the mechanical counterpart of the optical frequency comb.\cite{Kippenberg2018}

Much of the richer behavior shows up under internal resonance (IR),\cite{Antonio2012,PhysRevApplied.13.014049,10157971,Mouharrar2024} where two modes are brought to an integer frequency ratio, 1:2 or 1:3. The modes then trade energy efficiently,\cite{Chen2017,Zhang2025,Wu2026} and that exchange can destabilize into a range of bifurcations.\cite{PhysRevLett.121.244302} For graphene, gating away the out-of-plane symmetry is known to establish a 1:2 IR\cite{Asadi2021} and, above a threshold, to seed PFCs.\cite{Kekekler2022} The combs in those experiments grow out of Neimark--Sacker (secondary Hopf) bifurcations, so the underlying motion is quasi-periodic and the spectra come out broadband and incommensurate.\cite{PhysRevApplied.19.014031} Far less is known about the other possibility a single drive that locks the response onto pure integer harmonics in the manner of high-harmonic generation (HHG) in nonlinear optics\cite{RevModPhys.72.545,Takahashi2013} or about where period doubling\cite{PhysRevLett.47.1349,Mahboob2016} fits into 2D resonators at all. Reaching that regime is a question of balance: the quadratic and cubic terms must be set so that deterministic multi-wave mixing outruns the quasi-periodic instability.

Here we work in exactly that regime. A local gate sets the membrane tension, offsets the modal dispersion, and pulls the fundamental flexural mode onto a 1:2 IR with the higher mode at twice its frequency; the asymmetric potential that results carries a large effective quadratic nonlinearity.\cite{Eichler2013,PhysRevB.104.155434} A single drive tone is then enough to give a deterministic, rather than quasi-periodic, response. Integer harmonics switch on one above the other as the drive grows, and past a threshold they collapse into a dense, phase-locked comb followed by a period-doubling cascade.\cite{Feigenbaum1978,PhysRevLett.134.107201}

At still higher drive the comb does something less expected: its spacing doubles, the number of lines drops about twofold, and energy moves back into the even-order comb lines. We understand this reverse period-doubling step as a controlled exit from the chaotic comb, returning the resonator to a large-amplitude periodic motion. The same spectra fix the quadratic and cubic coefficients of the membrane. Because each of these regimes is reached on one device with nothing more than a change of gate or drive, the resonator serves as a switchable frequency multiplier\cite{HanWang2009} and comb source; the reverse bifurcation, in addition, points to a hardware route for bifurcation annealing in mechanical Ising machines.\cite{PhysRevResearch.4.013149}

\section*{Experimental Setup}

\begin{figure}
  \centering
  \includegraphics[width=\textwidth]{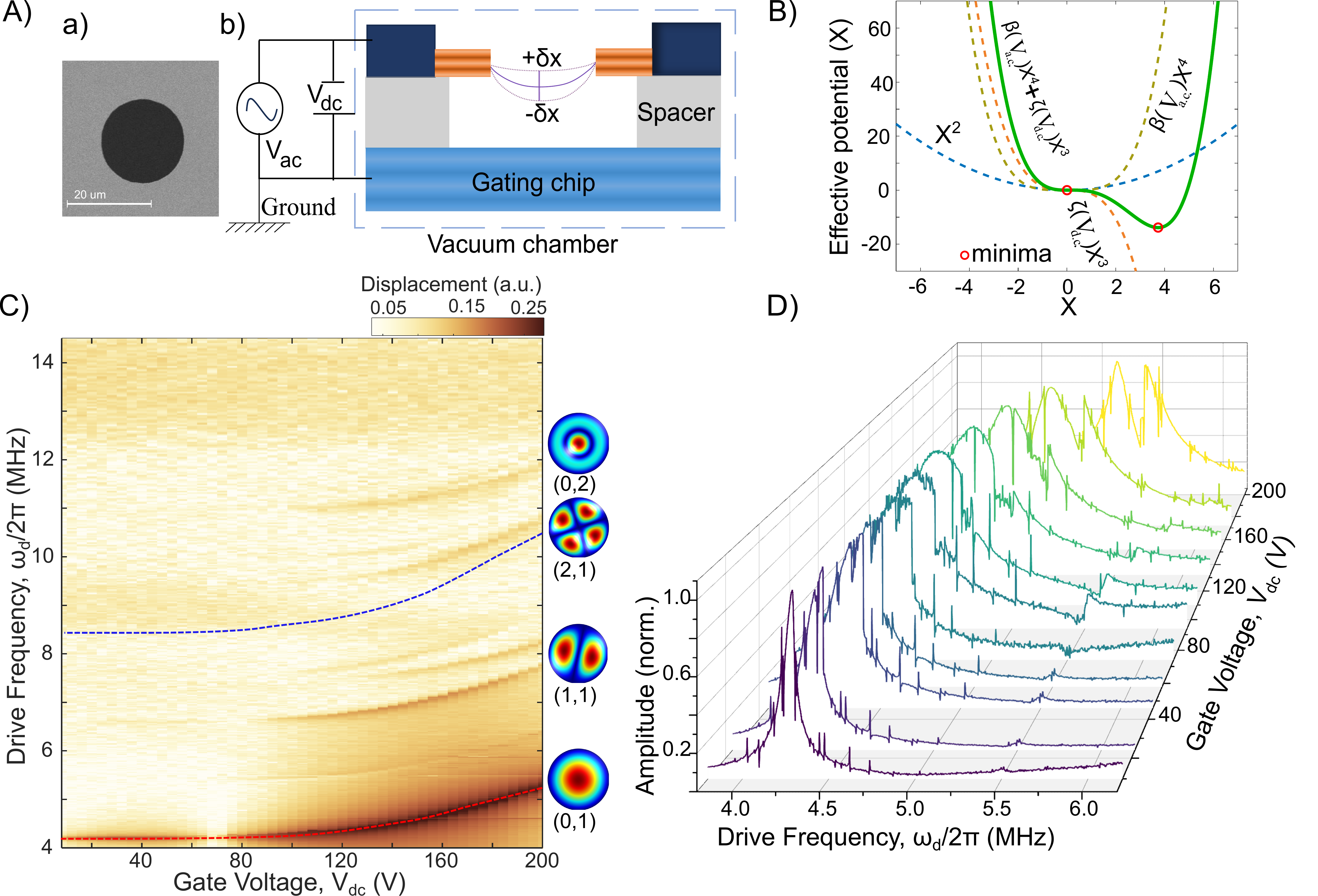}
  \caption{\textbf{Device, effective potential, and gate-tunable 1:2 internal resonance.} (A)(a)~Scanning electron micrograph of the device. (b)~Capacitive geometry: an AC voltage $\Vac$ drives the membrane and a DC gate voltage $\Vdc$ sets the static tension. (B)~Effective potential $V(X)$ of the fundamental mode, showing the competing quadratic and cubic nonlinear terms. (C)~Dispersion of the mechanical resonances versus $\Vdc$. (D)~Amplitude response of the fundamental mode versus $\Vdc$ at fixed $\Vac = 6.0$~V.}
  \label{fig:device}
\end{figure}

The device is a monolayer graphene drumhead about $20\, \mu$m in diameter, suspended over a circular hole etched in a silicon nitride (SiNx) membrane (Figure~\ref{fig:device}A(a)). The drum is contacted in a capacitive geometry (Figure~\ref{fig:device}A(b)): an AC voltage ($\Vac$) drives the membrane at a chosen frequency, while a DC gate voltage ($\Vdc$) sets its static tension. All measurements are done at room temperature in vacuum, with optical interferometric detection. The drive is supplied by a function generator and the spectral response is recorded on a spectrum analyzer.

Applying the gate voltage pulls the membrane toward the back gate and shifts its equilibrium position; the membrane then oscillates about this displaced position. Because the drum is curved, the path lengths for positive and negative out-of-plane displacements, $+\delta X$ and $-\delta X$, differ, so the tension is no longer symmetric in $X$. This geometric asymmetry breaks the out-of-plane symmetry of the membrane and adds a cubic-in-displacement term $\tfrac{1}{3}\zeta X^3$ to the potential (orange dashed line), alongside the symmetric elastic term $\tfrac{1}{2}K X^2$ (blue dashed line) and the quartic term $\tfrac{1}{4}\beta X^4$ that becomes important at large amplitude (yellow dashed line), as shown in Figure~\ref{fig:device}B. The total effective potential (green solid line) is
\begin{equation}
V(X) = \frac{1}{2}K X^{2} + \frac{1}{3}\zeta X^{3} + \frac{1}{4}\beta X^{4},
\label{eq:potential}
\end{equation}
where $K$ is the elastic constant, $\zeta$ the quadratic nonlinear coefficient, and $\beta$ the cubic (Duffing) coefficient. The degree of asymmetry is set by the gate voltage, which controls the equilibrium position and hence the nonlinear response.

Figure~\ref{fig:device}C shows the measured dispersion of the mechanical resonances against $\Vdc$, covering 4.0--14.5~MHz over a gate range of 8--200~V. These data were taken with a Zurich Instruments HF2LI lock-in amplifier at $\Vac = 1.4$~V, low enough to stay in the linear regime, where no Duffing response is detectable. At low $\Vdc$ the fundamental mode sits at $\omega_0/2\pi \approx 4.2102$~MHz (red dashed line) and a distinct higher-order mode appears at $\omega_2/2\pi$; the blue dashed line marks $2\omega_0/2\pi$. As $\Vdc$ increases, the electrostatic tension stiffens both modes, but $\omega_0/2\pi$ rises faster than $\omega_2/2\pi$. Near $\Vdc \approx 120$~V the two curves meet the condition $\omega_2 \approx 2\omega_0$, which sets up the 1:2 internal resonance.

The amplitude response of the fundamental mode confirms this picture. Sweeping $\Vdc$ at a fixed $\Vac = 6.0$~V (Figure~\ref{fig:device}D) gives a Duffing-like response with no splitting below $\Vdc \approx 120$~V. Above this voltage a split develops in the spectrum and deepens with increasing $\Vdc$. The split marks the onset of the internal resonance: the quadratic coupling becomes strong enough to transfer energy between the modes, producing an avoided-crossing like feature.

The 1:2 resonance is not a single-point coincidence. Once $\omega_2 \approx 2\omega_0$ is met, the two modes track one another, so the splitting survives across a broad gate window, $\Vdc \approx 150$--200~V. The harmonic, comb, and period-doubling data below were taken at $\Vdc = 158$, 172, and 179~V; these are simply convenient points inside that window and can be read on the same footing.

\section*{Integer Harmonic Generation}

\begin{figure}
  \centering
  \includegraphics[width=0.72\textwidth]{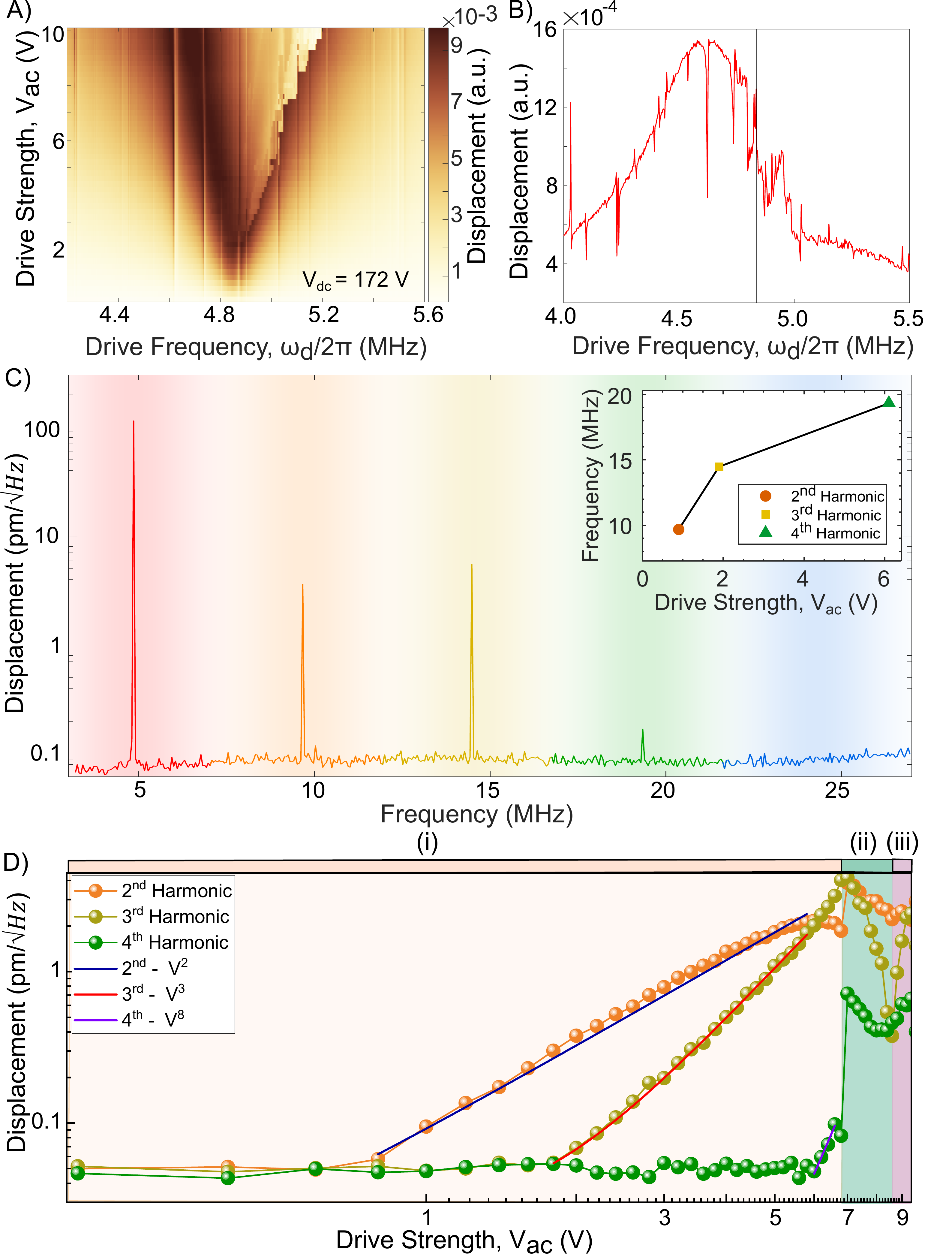}
  \caption{\textbf{Integer high-harmonic generation and amplitude scaling.} (A)~Fundamental-mode spectrum with increasing drive amplitude at $\Vdc = 172$~V; the spectral dip deepens as the nonlinear mode coupling strengthens. (B)~Spectrum at fixed drive amplitude, showing the mode splitting from internal resonance. The black vertical line marks the drive frequency used in the following measurements. (C)~Broadband spectrum under single-tone drive, showing integer harmonics up to fourth order ($n\omd$, $n = 1$--4). Inset: onset threshold of each harmonic versus drive amplitude. (D)~Amplitudes of the second, third, and fourth harmonics versus drive voltage, showing three regimes: (i)~integer-harmonic generation with $A_2 \propto V^2$, $A_3 \propto V^3$, $A_4 \propto V^8$; (ii)~comb formation; (iii)~reverse period-doubling. Solid lines are power-law fits in regime~(i).}
  \label{fig:harmonics}
\end{figure}

At gate voltages where $\omega_2 \approx 2\omega_0$, the fundamental mode undergoes second-harmonic generation, with two phonons from the fundamental feeding the higher-order mode. Increasing $\Vac$ at fixed $\Vdc$ first produces a dip in the resonance peak (Figure~\ref{fig:harmonics}A); the dip signals coupling to the higher-order mode near $2\omega_0$ and deepens as more energy is transferred. Figure~\ref{fig:harmonics}B shows the response at $\Vac = 8.3$~V, past the onset of the dip, with the vertical line marking the drive frequency.

Driving the fundamental mode with a single tone at $\omd/2\pi = 4.8338$~MHz generates integer harmonics. For $\Vac = 6.6$~V (Figure~\ref{fig:harmonics}C), harmonics up to fourth order ($n\omd/2\pi$, $n = 1$--4) appear, each above its own threshold. The inset shows that the second harmonic turns on first, with the third and fourth following at higher drive. This ordering reflects the competition between the two nonlinearities: the quadratic term $\zeta$ dominates at moderate drive, while the cubic term $\beta$ matters only at higher amplitude.

Figure~\ref{fig:harmonics}D plots the amplitudes of the second ($2\omd$), third ($3\omd$), and fourth ($4\omd$) harmonics against $\Vac$. In region~(i), the integer-harmonic regime, the low-drive data follow power laws $|A_2| \propto V_{\mathrm{ac}}^{2}$, $|A_3| \propto V_{\mathrm{ac}}^{3}$, and $|A_4| \propto V_{\mathrm{ac}}^{8}$. The quadratic scaling of $A_2$ identifies a $\zeta$-mediated three-wave-mixing process, in which two drive phonons combine, $\omd + \omd \to 2\omd$. The cubic scaling of $A_3$ has two contributions: direct generation through the Duffing term $\beta x^3$ at $3\omd$, and a cascaded $\zeta$-mediated step that sums the drive with the second harmonic, $\omd + 2\omd \to 3\omd$. Beyond a transition point the cascade dominates. The steep $V_{\mathrm{ac}}^{8}$ scaling of $A_4$ points to a higher-order process built from several cascaded $\zeta$ steps.

Above a threshold drive the system enters region~(ii) (Figure~\ref{fig:harmonics}D), where combs form around the drive frequency and its harmonics, with a comb spacing of $\Delta_c$. At the transition the second and fourth harmonics rise sharply while the third grows more gradually. Inside the comb regime the second and fourth harmonics transfer energy to their sidebands more slowly than the third. The matched behavior of the second and fourth harmonics suggests a shared underlying process, distinct from the one that governs the third.

Raising the drive further produces a second transition, a reverse period-doubling shift into region~(iii) (Figure~\ref{fig:harmonics}D). Here the number of comb lines drops to about half and the line spacing doubles to $\approx 2\Delta_c$, which returns energy to the harmonics so their amplitudes recover. The third harmonic shows the fastest loss and recovery, again setting it apart from the second and fourth.

\section*{Harmonic Comb Formation and Period-Doubling Dynamics}

\begin{figure}
  \centering
  \includegraphics[width=0.78\textwidth]{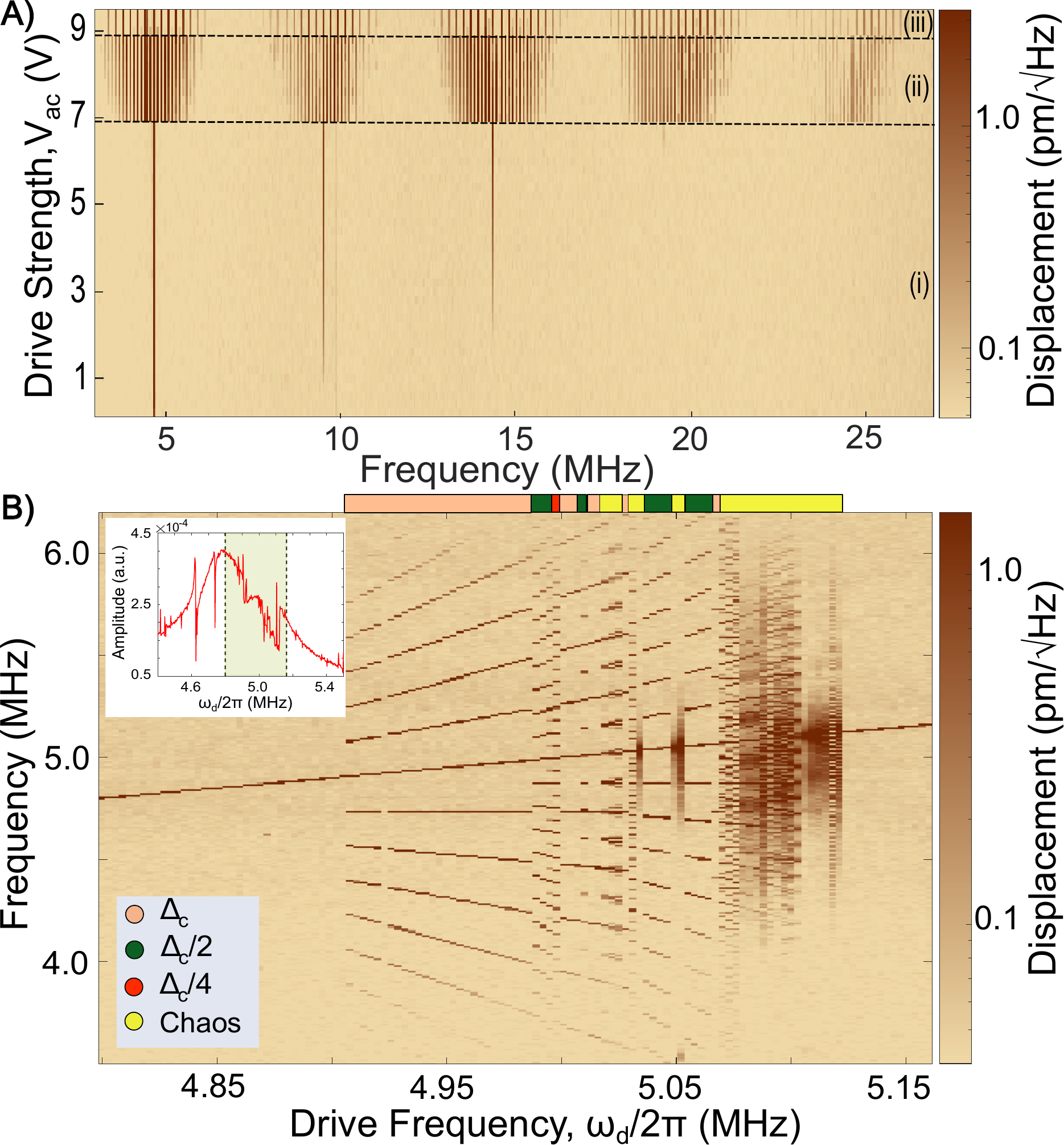}
  \caption{\textbf{From integer harmonics through frequency combs to chaos.} (A)~Spectral map versus drive amplitude at fixed $\omd/2\pi = 4.8338$~MHz and $\Vdc = 158$~V. Three regimes are marked: (i)~integer-harmonic generation ($\Vac \le 6.8$~V); (ii)~dense comb ($6.8 \le \Vac \le 8.8$~V); (iii)~reverse period-doubled comb with doubled spacing ($8.8 \le \Vac \le 9.4$~V). (B)~Spectral map versus drive frequency at fixed drive amplitude, showing the evolution from a single peak to a comb, then successive period-doubling steps ($\Delta_c \to \Delta_c/2 \to \Delta_c/4$) and finally chaos. Inset: fundamental-mode spectrum with the 1:2 mode splitting; the shaded band marks the swept frequency range.}
  \label{fig:combs}
\end{figure}

Figure~\ref{fig:combs}A is a two-dimensional spectral map versus drive amplitude $\Vac$ at fixed $\omd/2\pi = 4.8338$~MHz and $\Vdc = 158$~V. As $\Vac$ is raised past the integer-harmonic regime, the discrete harmonics give way first to a phononic frequency comb and then to a reverse period-doubled comb. The map shows three regimes separated by sharp thresholds.

In the integer-harmonic-generation (IHG) regime ($\Vac \le 6.8$~V), the spectrum holds discrete peaks at integer multiples of the drive ($\approx n\omd$), with the power-law amplitudes of Figure~\ref{fig:harmonics}D, region~(i). In the comb regime ($6.8 \le \Vac \le 8.8$~V), a dense set of equally spaced lines appears at multiples of the fundamental spacing $\Delta_c = 107.6$~kHz and extends across the measurement band. In the reverse period-doubling regime ($8.8 \le \Vac \le 9.4$~V), the spacing jumps from $\Delta_c$ to $2\Delta_c = 215.2$~kHz, the number of lines roughly halves, and the even-order comb lines regain amplitude, as in Figure~\ref{fig:harmonics}D, region~(iii).

\section*{Period-Doubling Cascade and Comb Spacing}

To map the bifurcation structure, we fix the drive amplitude and gate voltage ($\Vdc = 179$~V) and record the power spectrum while sweeping the drive frequency. Figure~\ref{fig:combs}B shows the resulting map versus $\omd/2\pi$ at $\Vac = 9.0$~V, swept from 4.80 to 5.16~MHz in steps of 3~kHz; the shaded band in the inset marks this range.

As $\omd/2\pi$ increases within the split-resonance region, the response changes from discrete harmonics to a dense comb, and the comb spacing $\Delta_c$ grows roughly linearly with drive frequency. Within this region, period doubling and reverse period doubling alternate. At $\omd/2\pi = 4.9860$~MHz a period-doubling bifurcation halves the spacing to $\Delta_c/2$; at 4.9950~MHz a second one quarters it to $\Delta_c/4$; beyond this point the spacing returns to $\Delta_c$, i.e., reverse period doubling.

These states are marked by the color bar at the top of Figure~\ref{fig:combs}B. At certain drive frequencies the successive halvings $\Delta_c \to \Delta_c/2 \to \Delta_c/4$ lead into chaos. This hierarchy of halved spacings is the spectral signature of a Feigenbaum period-doubling cascade. Although our frequency resolution is too coarse to extract the Feigenbaum constant $\delta \approx 4.669$, the observed sequence $\Delta_c$, $\Delta_c/2$, $\Delta_c/4$ followed by chaos matches the universal period-doubling route.

\section*{Discussion}

The measurements follow one graphene resonator through harmonic generation, comb formation, and a reverse period doubling, all under a gate-defined 1:2 internal resonance. The electrostatic gate plays a double role here: it cancels the modal dispersion that would otherwise spoil the commensurability, and it leaves the membrane sitting in an asymmetric potential whose quadratic term couples the fundamental strongly to the mode near $2\omega_0$.

What we do not see is the Neimark--Sacker quasi-periodicity that dominates the earlier graphene-comb experiments.\cite{Kekekler2022} The distinguishing observation here is a clean integer-harmonic regime that exhibits harmonics through the fourth order at high spectral purity, followed by comb formation, with the comb subsequently undergoing a period-doubling cascade.\cite{Feigenbaum1978} The harmonic regime can be switched on or off just by gating the dispersion in and out of the 1:2 condition.

Driven harder, the resonator bifurcates into the dense comb. The line spacing then evolves by period doubling and its reverse, the doubling to $2\Delta_c$ and the energy redistribution both setting in at sharp drive thresholds. The reverse branch is the useful one: it carries the system back out of the chaotic comb into stable, large-amplitude periodic motion. This diversity of dynamical regimes traces directly to the highly tunable nonlinear energy landscape of the graphene resonator.

Fitting the harmonic amplitudes fixes the membrane nonlinearities: a cubic (Duffing) coefficient $\beta \approx 1.01 \times 10^{20}$~N/m$^3$ and a quadratic coefficient $\zeta \approx 1.16 \times 10^{9}$~N/m$^2$ (Supporting Information). Both lie in the range expected for an atomically thin membrane, and the magnitude of $\zeta$ is the quantitative statement that the gate has driven the device into a strong quadratic, $\chi^{(2)}$-like, regime.

The practical appeal is that all of these behaviors live on one membrane and are selected by gate voltage alone. Held in the integer-harmonic regime, the device works as a zero-quiescent-power RF multiplier: drive at $\omd$, take the output at $n\omd$, and move the usable harmonic by re-gating rather than by rebuilding the resonator. The same device reaches a broadband comb and, through reverse period doubling, can be walked deliberately between comb line densities; tracking the harmonics likewise gives a handle for multi-spectral mass sensing. It is this controlled entry into and exit from a bifurcation that maps onto annealing in mechanical Ising machines.

 \section*{Acknowledgments}
This work is supported by the National Quantum Mission of DST, ERC Grant No.~639739, and DFG, German Research Foundation, projects 449506295 and 328545488, CRC/TRR 227. A.K.R. acknowledges MHRD for financial support. J.A.M. acknowledges the Prime Minister's Research Fellows (PMRF) scheme of the Ministry of Human Resource Development, Govt.\ of India, for financial support.
\section*{Supporting Information Available}
Additional spectral maps versus drive amplitude and drive frequency at other gate voltages; numerical simulations of the coupled-mode equations; the analytic single-mode-pair model under 1:2 internal resonance, including the symmetry-breaking origin of the intermodal coupling; and the extraction of the quadratic and cubic nonlinear coefficients.
%\bibliographystyle{naturemag} 
%\bibliographystyle{unsrt} 
%\bibliography{HHG}
%\bibliography{scibib}

%\onecolumn{
%	\bibliographystyle{apalike}
\bibliography{references}

\newpage

\title{Supporting Information: Tunable Nonlinear Landscapes in Graphene Nanoelectromechanical Systems}
\maketitle

\setcounter{equation}{0}
\renewcommand{\theequation}{S.\arabic{equation}}
\renewcommand\thefigure{S.\arabic{figure}}    
\setcounter{figure}{0} 
\makeatletter

\section*{Device}

\begin{figure}[!ht]
  \centering
  \includegraphics[width=0.45\textwidth]{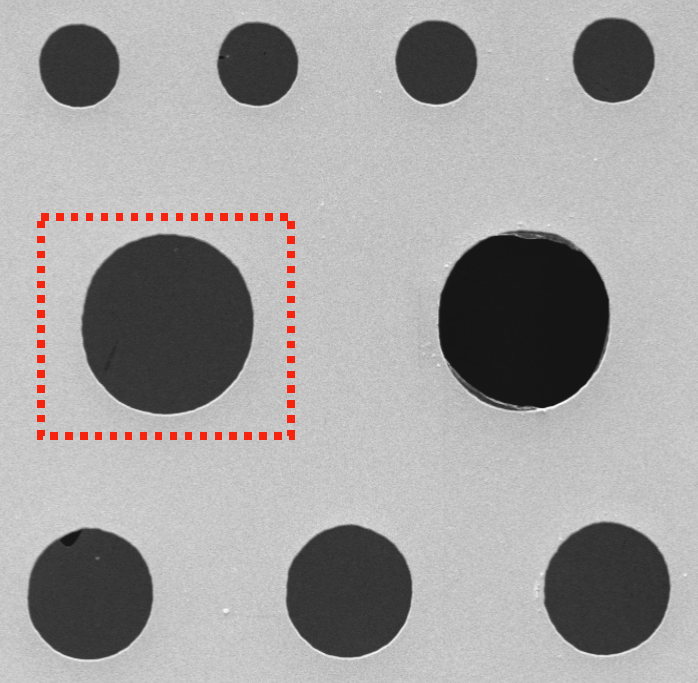}
  \caption{Scanning electron micrograph of the device.}
  \label{fig:sem}
\end{figure}

Figure~\ref{fig:sem} shows a scanning electron micrograph of a typical device: a monolayer graphene drum suspended over a circular hole in the silicon nitride (SiNx) membrane.

\section*{Spectral Map versus Drive Amplitude}

\begin{figure}[!ht]
  \centering
  \includegraphics[width=0.85\textwidth]{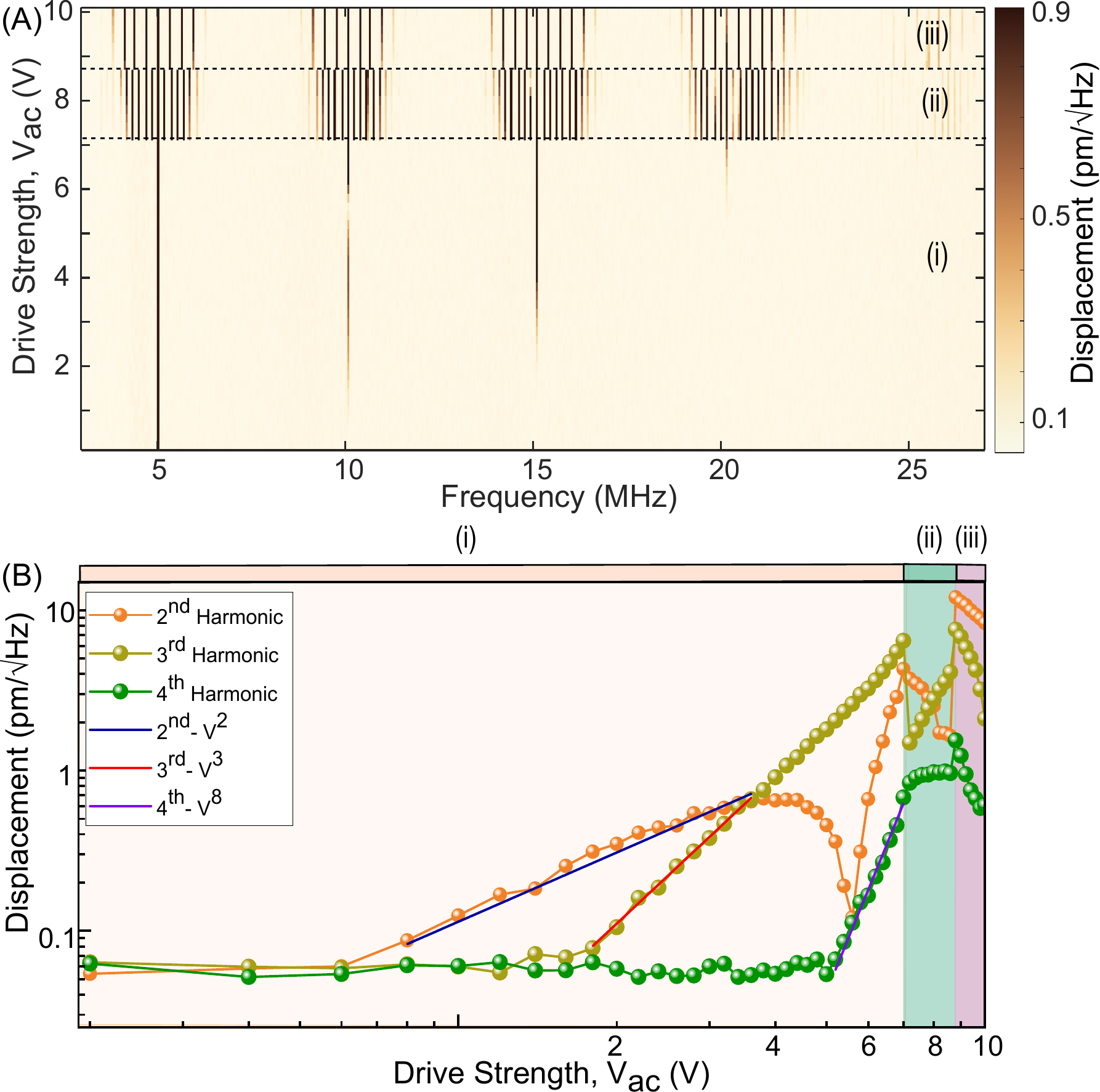}
  \caption{\textbf{Spectral evolution with drive amplitude.} (A)~Spectral map of the fundamental graphene mode under direct drive. With the mode at the 1:2 internal-resonance condition, the drive produces harmonic generation, then a frequency comb, and then reverse period-doubling. (B)~Peak amplitudes of the harmonic components. The second, third, and fourth harmonics scale as $V^2$, $V^3$, and $V^8$, matching the main text.}
  \label{fig:driveamp}
\end{figure}

Figure~\ref{fig:driveamp} repeats the drive-amplitude measurement of the main text at a different gate voltage. The fundamental mode is driven directly; at low drive the harmonics appear one at a time, each above its own threshold, and beyond a certain drive they broaden into a comb and then into the reverse period-doubling regime. The fitted peak amplitudes again scale as $V^2$, $V^3$, and $V^8$.

\section*{Spectral Map versus Drive Frequency}

\begin{figure}[!ht]
  \centering
  \includegraphics[width=0.9\textwidth]{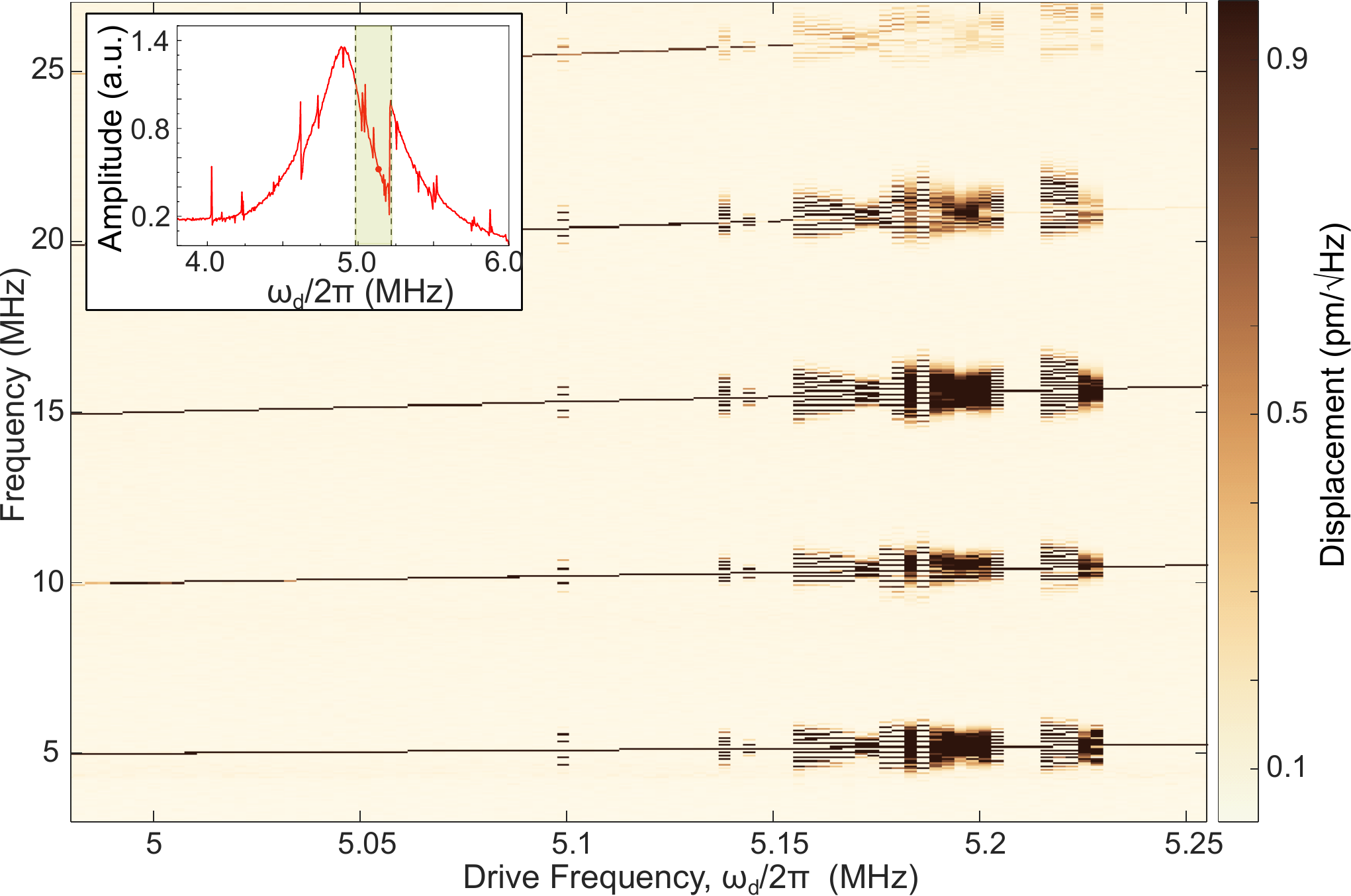}
  \caption{\textbf{Spectral evolution with detuning.} Spectral map of the nonlinear regimes versus drive frequency. The shaded band in the inset marks the frequency window, near the internal-resonance condition, over which the data were taken.}
  \label{fig:drivefreq}
\end{figure}

Figure~\ref{fig:drivefreq} is a second drive-frequency data set at a different gate voltage, tracking the first five harmonics as the drive frequency is swept at fixed amplitude. In the inset spectrum of the driven fundamental mode, the dip marks the 1:2 condition with the mode at roughly twice the frequency, and the shaded band marks the swept range.

\section*{Numerical Simulations}

\begin{figure}[htb]
  \centering
  \includegraphics[width=0.9\textwidth]{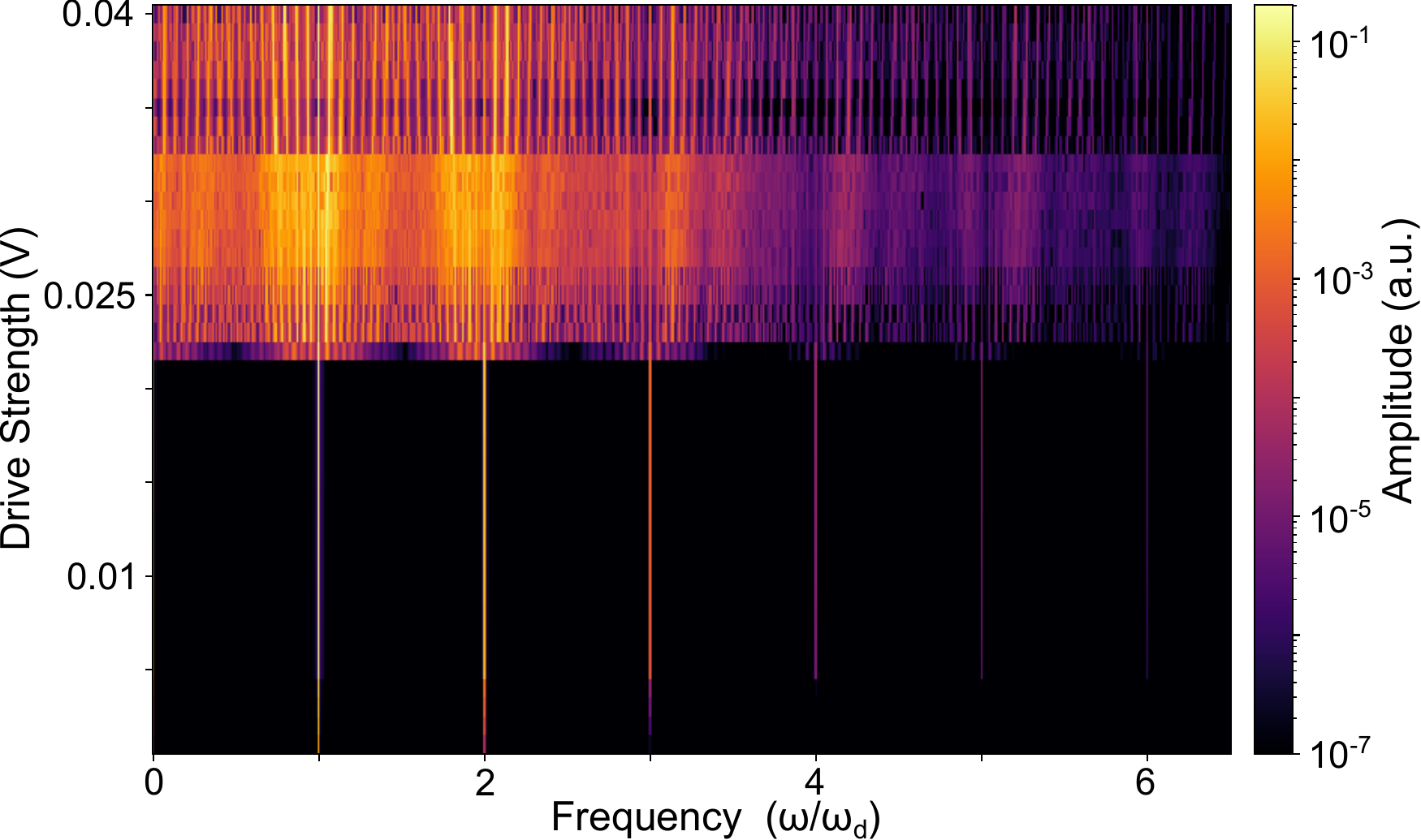}
  \caption{Simulated spectral map versus drive strength.}
  \label{fig:simamp}
\end{figure}

\begin{figure}[htb]
  \centering
  \includegraphics[width=0.9\textwidth]{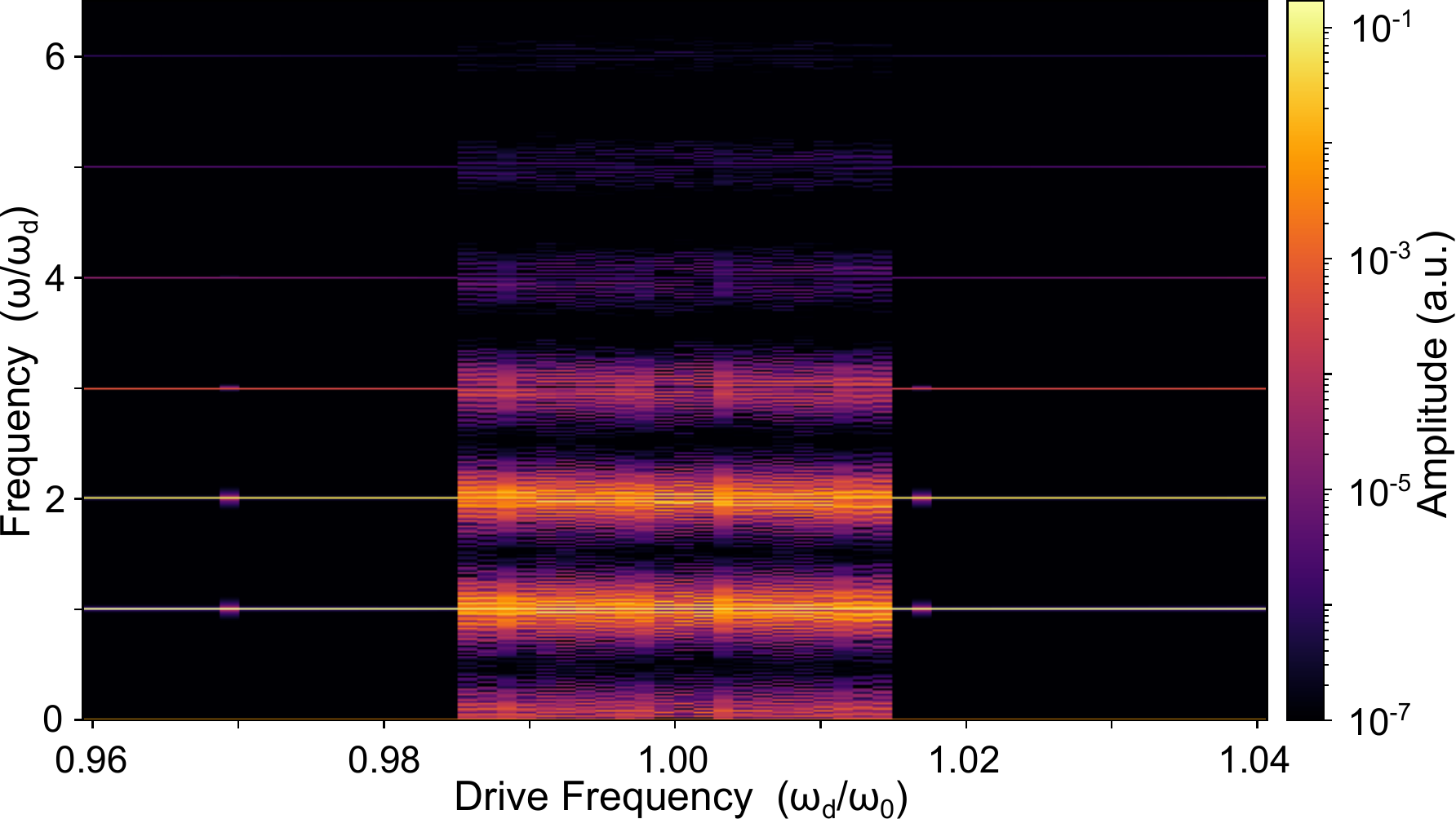}
  \caption{Simulated spectral map versus drive frequency.}
  \label{fig:simfreq}
\end{figure}

As a check against the data, we integrate the coupled-mode equations with a fourth-order Runge--Kutta scheme. In dimensionless form,
\begin{align}
\ddot{x}_1 + \gamma_1 \dot{x}_1 + \omega_1^2 x_1 + \beta_1 x_1^3 + 2\alpha x_1 x_2 &= F \cos(\omd t),
\label{eq:sim1}\\
\ddot{x}_2 + \gamma_2 \dot{x}_2 + \omega_2^2 x_2 + \beta_2 x_2^3 + \alpha x_1^2 &= 0,
\label{eq:sim2}
\end{align}
where $x_1$ is the fundamental-mode displacement, $x_2$ that of the higher-order mode, $\gamma_{1,2}$ are the damping rates, $\beta_{1,2}$ the Duffing coefficients, and $\alpha$ the quadratic intermodal coupling. The simulated drive-voltage map (Figure~\ref{fig:simamp}) reproduces the experimental progression: higher harmonics in turn, then a comb, then the more complex dynamics at high drive. The simulated drive-frequency map at fixed drive voltage (Figure~\ref{fig:simfreq}) likewise captures the main features of the measured maps.

\section*{Fundamental Mode under Internal Resonance}

\subsection*{Symmetry breaking and the intermodal coupling}

In the main text the broken out-of-plane symmetry adds a cubic term to the single-mode potential, $V(X) = \tfrac{1}{2}KX^2 + \tfrac{1}{3}\zeta X^3 + \tfrac{1}{4}\beta X^4$ (Eq.~(1)), so the restoring force $-\,\mathrm{d}V/\mathrm{d}X$ contains a quadratic part $\zeta X^2$. To describe the 1:2 internal resonance we expand the membrane displacement in the two participating eigenmodes,
\begin{equation}
X(\mathbf{r},t) = x_1(t)\,\phi_1(\mathbf{r}) + x_2(t)\,\phi_2(\mathbf{r}),
\label{eq:modal}
\end{equation}
where $\phi_1$ and $\phi_2$ are the (mass-normalized) mode shapes of the fundamental mode (frequency $\omega_1 \equiv \omega_0$) and the higher-order mode ($\omega_2 \approx 2\omega_1$), and $x_1$, $x_2$ are the modal amplitudes. Inserting Eq.~\eqref{eq:modal} into the quadratic force and expanding,
\begin{equation}
\zeta X^2 = \zeta\bigl(x_1\phi_1 + x_2\phi_2\bigr)^2
          = \zeta\bigl(\phi_1^2\, x_1^2 + 2\phi_1\phi_2\, x_1 x_2 + \phi_2^2\, x_2^2\bigr),
\label{eq:expand}
\end{equation}
the cross term $2\phi_1\phi_2\, x_1 x_2$ feeds the fundamental mode and the $\phi_1^2\, x_1^2$ term feeds the higher-order mode. Projecting onto each mode shape and writing the overlap-weighted coupling as
\begin{equation}
\alpha \equiv \zeta \int \phi_1^2\, \phi_2 \, \mathrm{d}A,
\label{eq:alpha}
\end{equation}
the equations of motion take the form of Eqs.~\eqref{eq:eom1} and \eqref{eq:eom2} below, with a $2\alpha x_1 x_2$ term in the fundamental-mode equation and an $\alpha x_1^2$ term in the higher mode. So $\alpha$ is just the main-text quadratic nonlinearity $\zeta$ weighted by the modal overlap. The integral in Eq.~\eqref{eq:alpha} is zero by symmetry for a flat membrane and turns on only when the gate breaks that symmetry, which is what makes the 1:2 resonance accessible.

\subsection*{Harmonic balance}

We model the system as two coupled modes under a 1:2 internal resonance,
\begin{align}
m_1 \ddot{x}_1 + c_1 \dot{x}_1 + k_1 x_1 + \beta x_1^3 + 2\alpha x_1 x_2 &= F \cos(\omd t),
\label{eq:eom1}\\
m_2 \ddot{x}_2 + c_2 \dot{x}_2 + k_2 x_2 + \alpha x_1^2 &= 0,
\label{eq:eom2}
\end{align}
where $m_i$, $c_i$, and $k_i = m_i \omega_i^2$ are the mass, damping, and spring constant of mode $i$ ($i = 1$ the fundamental, $i = 2$ the higher-order mode), $\beta$ is the Duffing coefficient of the driven fundamental mode, and $\alpha$ is the coupling of Eq.~\eqref{eq:alpha}. We seek the steady-state periodic response by single-term harmonic balance, taking
\begin{equation}
x_1(t) = X_1 \cos(\omd t + \varphi), \qquad x_2(t) = X_2 \cos\bigl(2(\omd t + \varphi)\bigr),
\label{eq:ansatz}
\end{equation}
where $X_1$ and $X_2$ are the steady-state amplitudes of the fundamental and higher-order modes and $\varphi$ is the phase of the response relative to the drive.

For the higher mode, Eq.~\eqref{eq:ansatz} in Eq.~\eqref{eq:eom2} with $x_1^2 = X_1^2 \cos^2(\omd t + \varphi) = \tfrac{1}{2} X_1^2 \bigl[1 + \cos\bigl(2(\omd t + \varphi)\bigr)\bigr]$ drives $x_2$ resonantly at $2\omd$ with amplitude $\tfrac{1}{2}\alpha X_1^2$. Balancing the $\cos\bigl(2(\omd t + \varphi)\bigr)$ components and dropping the off-resonant damping term, $\bigl(k_2 - 4 m_2 \omd^2\bigr) X_2 + \tfrac{1}{2}\alpha X_1^2 = 0$, so
\begin{equation}
X_2 = \frac{\alpha X_1^2}{2 m_2 \bigl(4\omd^2 - \omega_2^2\bigr)}, \qquad \omega_2^2 = \frac{k_2}{m_2}.
\label{eq:X2}
\end{equation}

\begin{figure}[!ht]
  \centering
  \includegraphics[width=0.7\textwidth]{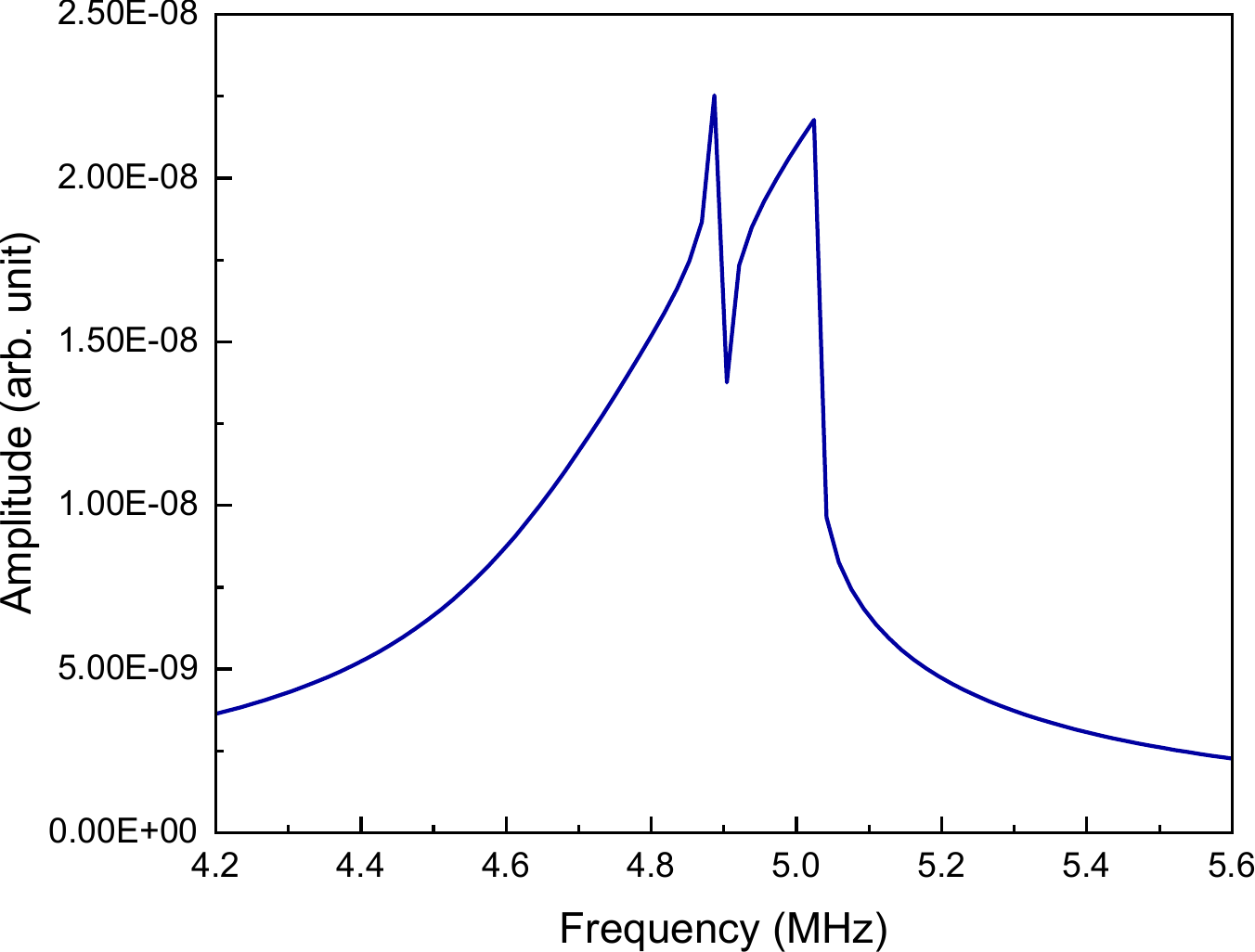}
  \caption{\textbf{Mode response under internal resonance.} Amplitude response of the fundamental mode coupled to the higher-order mode, from the positive real roots of Eq.~\eqref{eq:sextic}.}
  \label{fig:iranalytic}
\end{figure}

For the fundamental, Eq.~\eqref{eq:ansatz} in Eq.~\eqref{eq:eom1} with the resonant parts of the nonlinear terms,
\begin{align}
x_1^3 &= X_1^3 \Bigl[\tfrac{3}{4}\cos(\omd t + \varphi) + \tfrac{1}{4}\cos\bigl(3(\omd t + \varphi)\bigr)\Bigr],
\label{eq:cube}\\
x_1 x_2 &= X_1 X_2 \cos(\omd t + \varphi)\cos\bigl(2(\omd t + \varphi)\bigr)
         = \tfrac{1}{2} X_1 X_2 \Bigl[\cos(\omd t + \varphi) + \cos\bigl(3(\omd t + \varphi)\bigr)\Bigr],
\label{eq:cross}
\end{align}
we keep only the components at $\omd$. Resolving the drive $F\cos(\omd t)$ along the in-phase and quadrature directions of the response gives the two balances
\begin{align}
m_1\bigl(\omega_1^2 - \omd^2\bigr) X_1 + \tfrac{3}{4}\beta X_1^3 + \alpha X_1 X_2 &= F \cos\varphi \quad \text{(in-phase)},
\label{eq:inphase}\\
c_1 \omd X_1 &= F \sin\varphi \quad \text{(quadrature)}.
\label{eq:quad}
\end{align}
Squaring and adding the two balances eliminates the drive phase $\varphi$ and yields the amplitude equation
\begin{equation}
\Bigl[ m_1\bigl(\omega_1^2 - \omd^2\bigr) + \alpha X_2 + \tfrac{3}{4}\beta X_1^2 \Bigr]^2 X_1^2
+ c_1^2 \omd^2 X_1^2 = F^2.
\label{eq:ampeq}
\end{equation}
Substituting $X_2$ from Eq.~\eqref{eq:X2} into Eq.~\eqref{eq:ampeq},
\begin{equation}
\Biggl[ m_1\bigl(\omega_1^2 - \omd^2\bigr)
+ \frac{\alpha^2 X_1^2}{2 m_2 \bigl(4\omd^2 - \omega_2^2\bigr)}
+ \tfrac{3}{4}\beta X_1^2 \Biggr]^2 X_1^2
+ c_1^2 \omd^2 X_1^2 = F^2.
\label{eq:ampeq2}
\end{equation}
Introducing the effective nonlinearity and detuning,
\begin{equation}
\beta_{\mathrm{eff}} = \frac{\alpha^2}{2 m_2 \bigl(4\omd^2 - \omega_2^2\bigr)} + \frac{3}{4}\beta,
\qquad
\Delta = \omega_1^2 - \omd^2,
\label{eq:betaeff}
\end{equation}
so that $\alpha X_2 + \tfrac{3}{4}\beta X_1^2 = \beta_{\mathrm{eff}} X_1^2$, Eq.~\eqref{eq:ampeq} reduces to the sextic in $X_1$,
\begin{equation}
\beta_{\mathrm{eff}}^2\, X_1^6 + 2 m_1 \Delta\, \beta_{\mathrm{eff}}\, X_1^4
+ \Bigl[ m_1^2 \Delta^2 + \bigl(c_1 \omd\bigr)^2 \Bigr] X_1^2 - F^2 = 0.
\label{eq:sextic}
\end{equation}
The steady-state amplitudes are the positive real roots of $X_1^2$ in Eq.~\eqref{eq:sextic}, plotted in Figure~\ref{fig:iranalytic}. The first term of $\beta_{\mathrm{eff}}$ in Eq.~\eqref{eq:betaeff} diverges as $\omega_2 \to 2\omd$, so close to the 1:2 resonance this coupling-induced term overwhelms the bare Duffing nonlinearity; that dominance is what the main text means by the enhanced quadratic response.

\section*{Extraction of the Nonlinear Coefficients}

The single-mode coefficients below use the main-text notation: $m$ is the effective mass, $\gamma$ the damping rate ($c_1 = m\gamma$), $\omega_0$ the fundamental frequency ($\omega_0 \equiv \omega_1$), and $A_n$ the measured complex amplitude at $n\omd$.

\subsection*{Quadratic Nonlinearity $\zeta$}

We start from the equation of motion with a quadratic term,
\begin{equation}
m\ddot{x} + m\gamma \dot{x} + m\omega_0^2 x + \zeta x^2 = F \cos(\omd t).
\label{eq:eomquad}
\end{equation}
A quadratic nonlinearity drives a response at twice the drive frequency, $2\omd$. Keeping the fundamental and second-harmonic components,
\begin{equation}
x(t) = A_1 e^{-i\omd t} + A_2 e^{-i 2\omd t} + \mathrm{c.c.},
\label{eq:ansatz2}
\end{equation}
where $A_1$ and $A_2$ are the amplitudes at $\omd$ and $2\omd$. The quadratic term produces a $2\omd$ component proportional to $A_1^2$. Balancing terms at $2\omd$,
\begin{equation}
A_2 = \frac{\zeta A_1^2}{m\bigl(4\omd^2 - \omega_0^2 + i\,2\gamma\omd\bigr)},
\label{eq:A2}
\end{equation}
and taking the magnitude,
\begin{equation}
|A_2| = \frac{|\zeta|\,|A_1|^2}{m\sqrt{\bigl(4\omd^2 - \omega_0^2\bigr)^2 + \bigl(2\gamma\omd\bigr)^2}}.
\label{eq:A2mag}
\end{equation}
The quadratic coefficient is therefore
\begin{equation}
\zeta = \frac{m\,|A_2|}{|A_1|^2}\,\sqrt{\bigl(4\omd^2 - \omega_0^2\bigr)^2 + \bigl(2\gamma\omd\bigr)^2}.
\label{eq:zeta}
\end{equation}
Substituting the measured amplitudes and parameters gives $\zeta = 1.16 \times 10^{9}$~N/m$^2$.

\subsection*{Cubic Nonlinearity $\beta$}

For the cubic term, the equation of motion is
\begin{equation}
m\ddot{x} + m\gamma \dot{x} + m\omega_0^2 x + \beta x^3 = F \cos(\omd t).
\label{eq:eomcubic}
\end{equation}
A cubic nonlinearity drives a response at $3\omd$. Keeping the fundamental and third-harmonic components,
\begin{equation}
x(t) = A_1 e^{-i\omd t} + A_3 e^{-i 3\omd t} + \mathrm{c.c.},
\label{eq:ansatz3}
\end{equation}
where $A_1$ and $A_3$ are the amplitudes at $\omd$ and $3\omd$. To leading order the cubic term gives a $3\omd$ component proportional to $A_1^3$ (the remaining resonant piece, $3|A_1|^2 A_1$, oscillates at $\omd$ and renormalizes the fundamental response). Balancing terms at $3\omd$,
\begin{equation}
A_3 = \frac{\beta A_1^3}{m\bigl(9\omd^2 - \omega_0^2 + i\,3\gamma\omd\bigr)},
\label{eq:A3}
\end{equation}
and taking the magnitude,
\begin{equation}
|A_3| = \frac{|\beta|\,|A_1|^3}{m\sqrt{\bigl(9\omd^2 - \omega_0^2\bigr)^2 + \bigl(3\gamma\omd\bigr)^2}}.
\label{eq:A3mag}
\end{equation}
The cubic coefficient is therefore
\begin{equation}
\beta = \frac{m\,|A_3|}{|A_1|^3}\,\sqrt{\bigl(9\omd^2 - \omega_0^2\bigr)^2 + \bigl(3\gamma\omd\bigr)^2}.
\label{eq:beta}
\end{equation}
Substituting the measured amplitudes and parameters gives $\beta = 1.01 \times 10^{20}$~N/m$^3$.

\begin{figure}[!ht]
  \centering
  \includegraphics[width=0.9\textwidth]{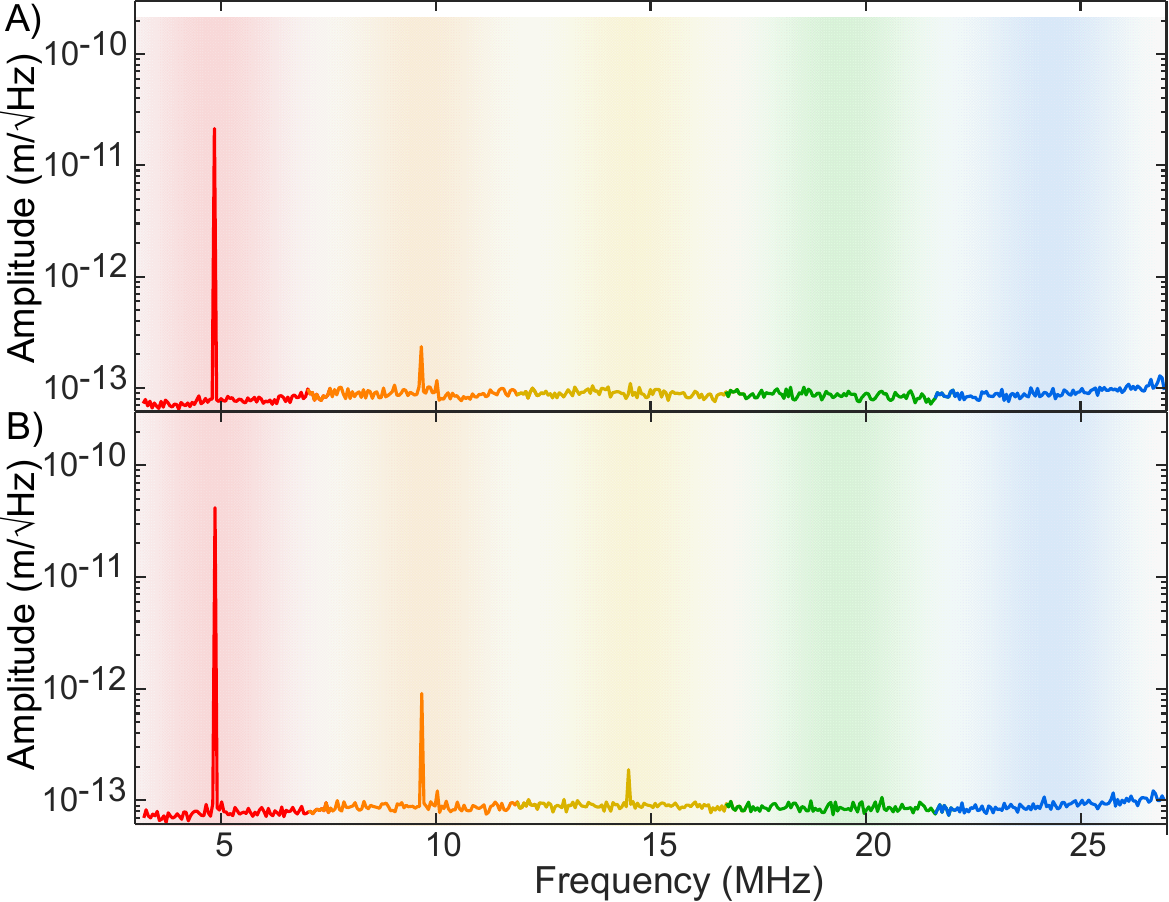}
  \caption{\textbf{High-harmonic generation.} (A)~Spectrum at the onset of the second harmonic. (B)~Spectrum once the third harmonic is also active.}
  \label{fig:hhgspectra}
\end{figure}

%\bibliography{HHG}

\end{document}